\documentclass[conference]{IEEEtran}
\usepackage{cite}
\usepackage{amsmath,amssymb,amsfonts}
\usepackage{algorithmic}
\usepackage{graphicx}
\usepackage{textcomp}
\usepackage{xcolor}
\usepackage{glossaries}
\usepackage{booktabs}
\usepackage{verbatim}
\usepackage{subcaption}
\usepackage[normalem]{ulem} 
\usepackage{multirow}
\newcommand{\splitcell}[2][c]{%
  \begin{tabular}{@{}#1@{}}
  #2
  \end{tabular}%
}

\usepackage[obeyFinal]{todonotes}

\makeatletter
\newcommand{\linebreakand}{%
  \end{@IEEEauthorhalign}
  \hfill\mbox{}\par
  \mbox{}\hfill\begin{@IEEEauthorhalign}
}
\makeatother

\newcommand{\PP}{\mathbf{P}}
\newcommand{\f}{\mathbf{f}}
\newcommand{\x}{\mathbf{x}}

\usepackage{pifont}
\newcommand{\cmark}{\ding{51}}%
\newcommand{\xmark}{\ding{55}}%

\newacronym{cnn}{CNN}{Convolutional Neural Network}
\newacronym{mfcc}{MFCC}{Mel-Frequency Cepstral Coefficient}
\newacronym{stft}{STFT}{Short-Time Fourier Transform}
\newacronym{imfcc}{IMFCC}{Inverted MFCC}
\newacronym{gsv}{GSV}{Gaussian Supervector}
\newacronym{gmm}{GMM}{Gaussian Mixture Model}
\newacronym{svm}{SVM}{Support Vector Machine}
\newacronym{psd}{PSD}{Power Spectral Density}
\newacronym{vad}{VAD}{Voice Activity Detection}
\newacronym{awgn}{AWGN}{Additive White Gaussian Noise}
\newacronym{dncnn}{DnCNN}{Denoising CNN}
\newacronym{snr}{SNR}{Signal-to-Noise Ratio}
\newacronym{rasta-mfcc}{RASTA-MFCC}{RASTA filtered Mel-Frequency Cepstral Coefficients}
\newacronym{bed}{BED}{Band Energy Difference}
\newacronym{rasta}{RASTA}{RelAtive Spectral TrAnsform}
\newacronym{htk}{HTK}{Hidden Markov Model Toolkit}
\newacronym{rbf}{RBF}{Radial Basis Function}

\def\BibTeX{{\rm B\kern-.05em{\sc i\kern-.025em b}\kern-.08em
    T\kern-.1667em\lower.7ex\hbox{E}\kern-.125emX}}
    
\begin{document}
\title{Speaker-Independent Microphone Identification \\ in Noisy Conditions}

\author{\IEEEauthorblockN{Antonio Giganti\IEEEauthorrefmark{1},
Luca Cuccovillo\IEEEauthorrefmark{2},
Paolo Bestagini\IEEEauthorrefmark{1},
Patrick Aichroth\IEEEauthorrefmark{2},
Stefano Tubaro\IEEEauthorrefmark{1}
}
\IEEEauthorblockA{\IEEEauthorrefmark{1}\textit{Dipartimento di Elettronica, Informazione e Bioingegneria - Politecnico di Milano - Milan, Italy}}
\IEEEauthorblockA{\IEEEauthorrefmark{2}\textit{Fraunhofer Institute for Digital Media Technology - TU Ilmenau - Ilmenau, Germany}}
\IEEEauthorblockA{\{antonio.giganti, paolo.bestagini, stefano.tubaro\}@polimi.it, \{luca.cuccovillo, patrick.aichroth\}@idmt.fraunhofer.de}
}

\maketitle


\begin{abstract}
This work proposes a method for source device identification from speech recordings that applies neural-network-based denoising, to mitigate the impact of counter-forensics attacks using noise injection. The method is evaluated by comparing the impact of denoising on three state-of-the-art features for microphone classification, determining their discriminating power with and without denoising being applied.
The proposed framework achieves a significant performance increase for noisy material, and more generally, validates the usefulness of applying denoising prior to device identification for noisy recordings.
\end{abstract}

\section{Introduction}
Speech recordings often contain relevant information for legal trials, and therefore represent potential evidence. In order to be used, however, they need to be checked for possible manipulations, which requires the development of reliable forensic analysis techniques, including source device identification, which aims at determining which recording device was used to capture a given recording. However, while source device identification approaches have become increasingly effective over the years, they suffer from being vulnerable to counter-forensics attacks which are based on introducing noise into the material. To address this problem, we propose and evaluate denoising using a neural network based approach: We provide an overview over relevant device identification approaches and the aforementioned vulnerability in Sec. \ref{section:sota}, propose a suitable denoising-based approach in Sec. \ref{section:method}, and evaluate it in Sec. \ref{section:evaluation}.

\section{State-of-the-Art}
\label{section:sota}

Several methods for device identification extract \glspl{mfcc} from input speech utterances~\cite{hanilci_2012, kotropoulos_2014}, and apply supervised classifiers to identify the respective devices. To improve performance, many researchers focused on \emph{removing} all information that they considered irrelevant for device identification, such as the speech content itself.

For instance, \cite{aggarwal_2014} constructs a device fingerprint by first extracting \glspl{mfcc} from speech-free frames, then training individual \gls{gmm} for each device. A \gls{gsv} is then built by concatenating statistics from the \gls{gmm} components to construct a template for each device, then to be used for identification using a \gls{svm} classifier. \glspl{gsv} are also used in~\cite{jiang_2019}, to employ a kernel-based projection method to map raw \gls{gsv} onto another dimensional space, in which microphone and speech signal information can be separated more easily.

In~\cite{hanilci_2014, pandey_2014}, an estimation of the \gls{psd} from a speech-free region of the audio recordings is computed to characterize the recording device. Both approaches use \gls{vad} to detect the silence parts in the audio track. In~\cite{shen_2021}, however, spectral subtraction of the speech component is used as an alternative to the \gls{vad}: After estimating the inherent noise of the device, they compute the \glspl{mfcc} to perform the identification task.

In \cite{luo_2018}, \textit{\gls{bed}} is proposed as a new audio feature for device identification. It relies on deviations between adjacent bands, which exist consistent across multiple devices and recording conditions, including different speakers and recording environments. Similarities between frequency bands were also exploited in~\cite{lin_2020}: Using attention models for spectrograms of noiseless and noisy audio recordings, authors found that low and high frequency ranges were most relevant for identifying devices from the same manufacturer. In addition, ~\cite{qin_2018, baldini_2019} found out that mid frequencies are most relevant for distinguishing devices of different manufacturers. Lastly, ~\cite{cuccovillo_2013,cuccovillo_2016} propose to estimate the frequency response directly using a spectrum classification approach, and use the entire spectrum of the estimated responses as classification feature.

While being very effective for noiseless material, the aforementioned methods did not yet investigate and address the effects of adding noise in much detail. However, noise addition causes masking of the devices' signatures in different parts of the spectrum, and opens the door for a simple counter-forensics attacks. Indeed, \gls{awgn} can be added to the recording by an attacker to simulate the attenuation due to path loss~\cite{baldini_2019}, or to obfuscate the device fingerprint and prevent identification completely~\cite{ho_handbook_2015}.

\section{Proposed Method}
\label{section:method}

Our proposed approach aims at addressing the aforementioned vulnerability with respect to noise addition, and increase the classification robustness for noisy material at the same time. For this purpose, we introduce a pre-processing step for Gaussian denoising within the time-frequency domain, and suggest the use of a neural network architecture for denoising to achieve an optimal balance: Too aggressive denoising can potentially destroy relevant device information, while too weak denoising is potentially ineffective with respect to robustness against noise.

In \cite{cuccovillo_2022_arXiv}, we focused on determining whether and to which extent a denoiser can enhance microphone classification. In this paper, we build on the respective results to determine whether denoising is beneficial not only to one specific feature, but to several key features for microphone classification. To achieve this, we perform an in-depth comparison between three different fingerprints for microphone classification and evaluate them in several noisy conditions, considering the influence of denoising on all three of them. Furthermore, we investigate the effect of denoising within the classifier training phase, measuring performance effects of data-driven denoising within training and testing.

We formulate the described device identification challenge for a closed-set scenario, i.e. considering a finite number of possible devices. More formally, given a set $\mathbf{\mathcal{X}}=\left\{\x_1, \x_2, \ldots, \x_I\right\}$ of $I$ audio recordings and a set $\mathcal{C}=\left\{c_1, c_2, \ldots, c_J\right\}$ of $J$ possible device models, our goal is to associate $\x_i \in \mathcal{X}$ to a specific $c \in \mathcal{C}$ producing the pair $(\x_i, c_j)$, indicating that the $i$-th recording was recorded using the $j$-th device. The identification task relies on estimating a particular fingerprint from a selected device -- in our case describing its microphone and audio acquisition pipeline -- and using it to differentiate between different models. In particular, we want to show that it is possible to introduce a denoising step into a classical classification pipeline to increase the robustness of device identification techniques to the presence of adversarial noise.

\begin{figure}[ht]
    \centering
    \includegraphics[width=.8\columnwidth]{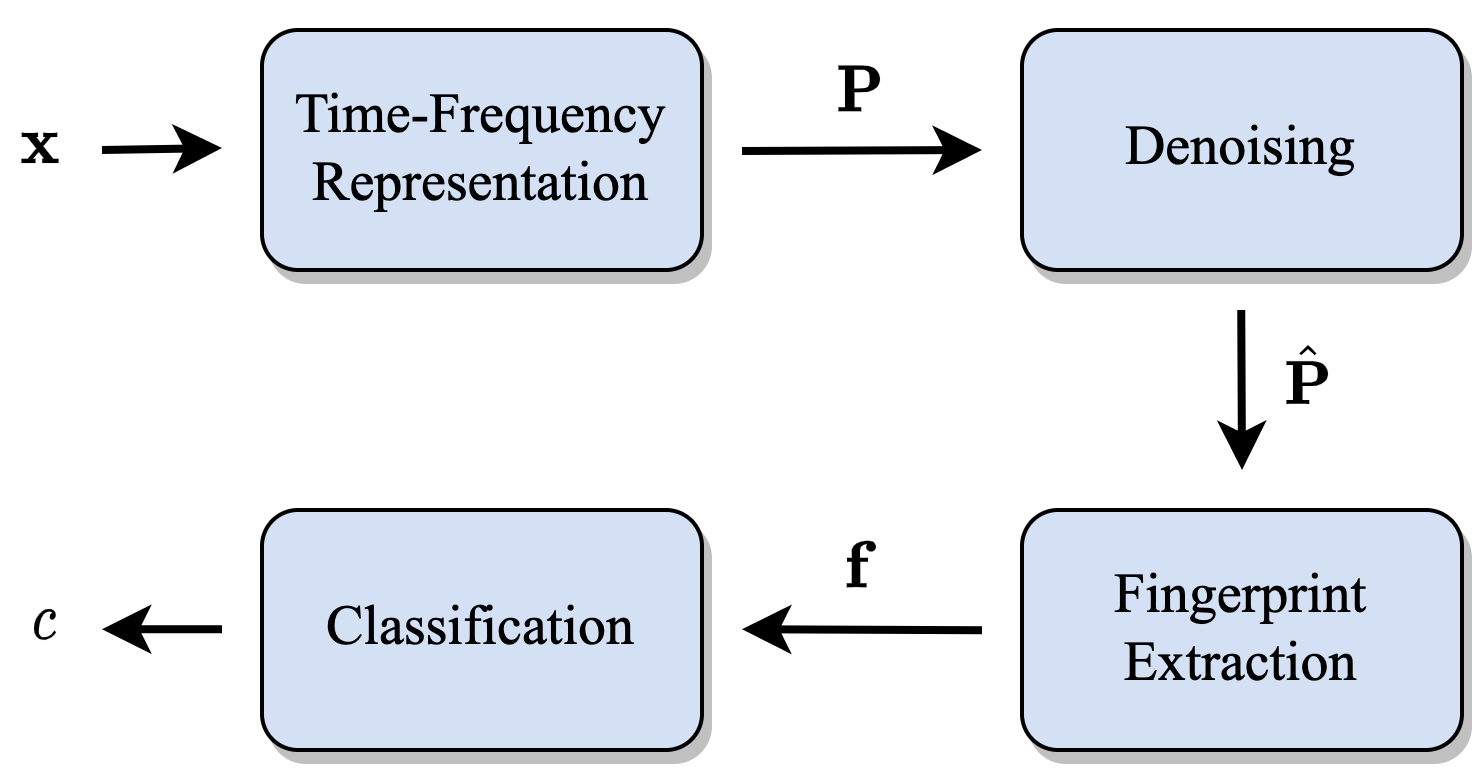}
    \caption{The workflow of our proposed method. Signal $\x$ is mapped to the time-frequency domain, cleaned up via denoising and then used for feature extraction and classification.}
    \label{fig:dummy proposed method flowchart}
\end{figure}

Figure~\ref{fig:dummy proposed method flowchart} depicts a general scheme of our proposed approach. In the following, we will describe the functionality of each block, and explain the motivations behind the respective design choices.

\subsection{Time-Frequency Representation}
Starting from an audio recording $\x$ in the time domain with a sampling rate of $F_\text{s}$ Hz, we calculate its time-frequency representation through the \gls{stft}. We adopt an analysis window of $N$ samples length and map the resulting spectrogram to a decibel scale, obtaining a so-called Log-Distributed Power Spectrogram
\begin{equation*}
    \PP = 10\log|\operatorname{STFT}(\x)|^2,
\end{equation*}
where $\operatorname{STFT}(\cdot)$ indicates the time-frequency transformation performed by the \gls{stft}, and $\lvert\cdot\rvert$ its magnitude.

\subsection{Denoising} 
By using the time-frequency representation of the audio signal, we can take advantage of well-established methods for image denoising.
We propose to first split the time-frequency representation $\PP$ into $K$ non-overlapping patches with fixed size $M\times M$, each of which may contain both noise and speech components. The number of patches depends on the length of the recording under analysis and does not affect the method, as long as at least one patch can be extracted. To reduce the computation requirements, we set $M=N/2$ and let each patch describe only the positive half of the spectrum.

Given the $k$-th patch $\PP_k$, we then propose to use the \gls{dncnn}~\cite{zhang_2017} neural network to perform a blind Gaussian noise removal in the time-frequency domain, which according to our findings in \cite{cuccovillo_2022_arXiv}, represents the currently best performing approach for denoising the spectrum while preserving microphone characteristics. More formally, we compute each denoised spectrogram patch as
\begin{equation*}
    \hat{\PP}_k = \operatorname{DEN}(\PP_k) = \PP_k - \operatorname{DnCNN}(\PP_k),
\end{equation*}
where $\operatorname{DEN}(\cdot)$ denotes the denoising operator, $\operatorname{DnCNN}(\cdot)$ stands for the noise estimation operation carried out by our trained network, and $\hat{\PP}_k$ is the denoised version of $\PP_k$ obtained by noise subtraction.

Lastly, we aggregate all denoised patches $\hat{\PP}_k$ into a single denoised spectrogram $\hat{\PP}$ to build the final denoising output.

\begin{figure}[ht]
    \centering
    \begin{subfigure}{0.35\columnwidth}
          \centering
          \includegraphics[width=1\columnwidth]{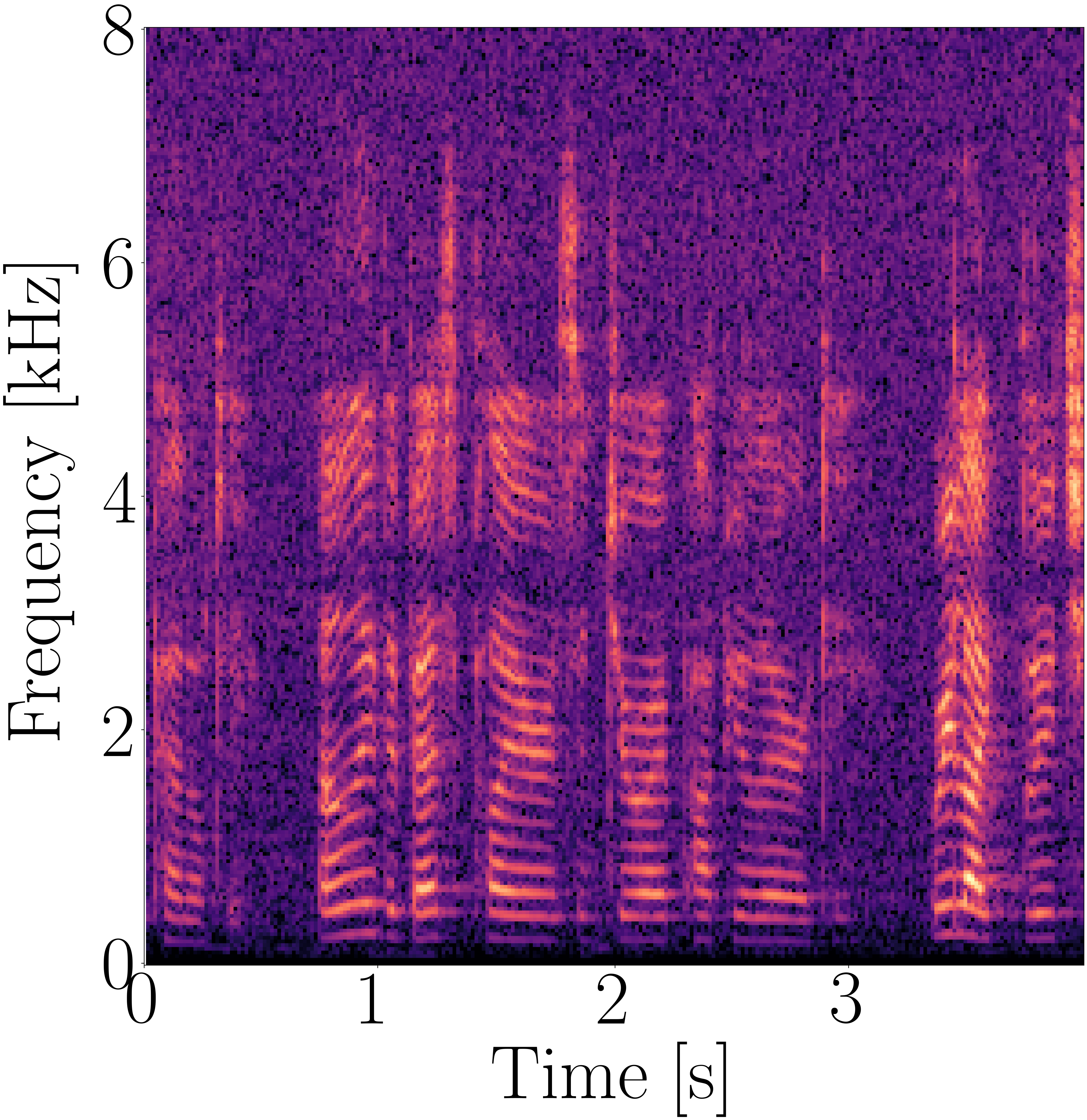}  
          \caption{Noisy.}
          \label{subfig:noisy}
    \end{subfigure}
    \begin{subfigure}{0.31\columnwidth}
          \centering
          \includegraphics[width=1\columnwidth]{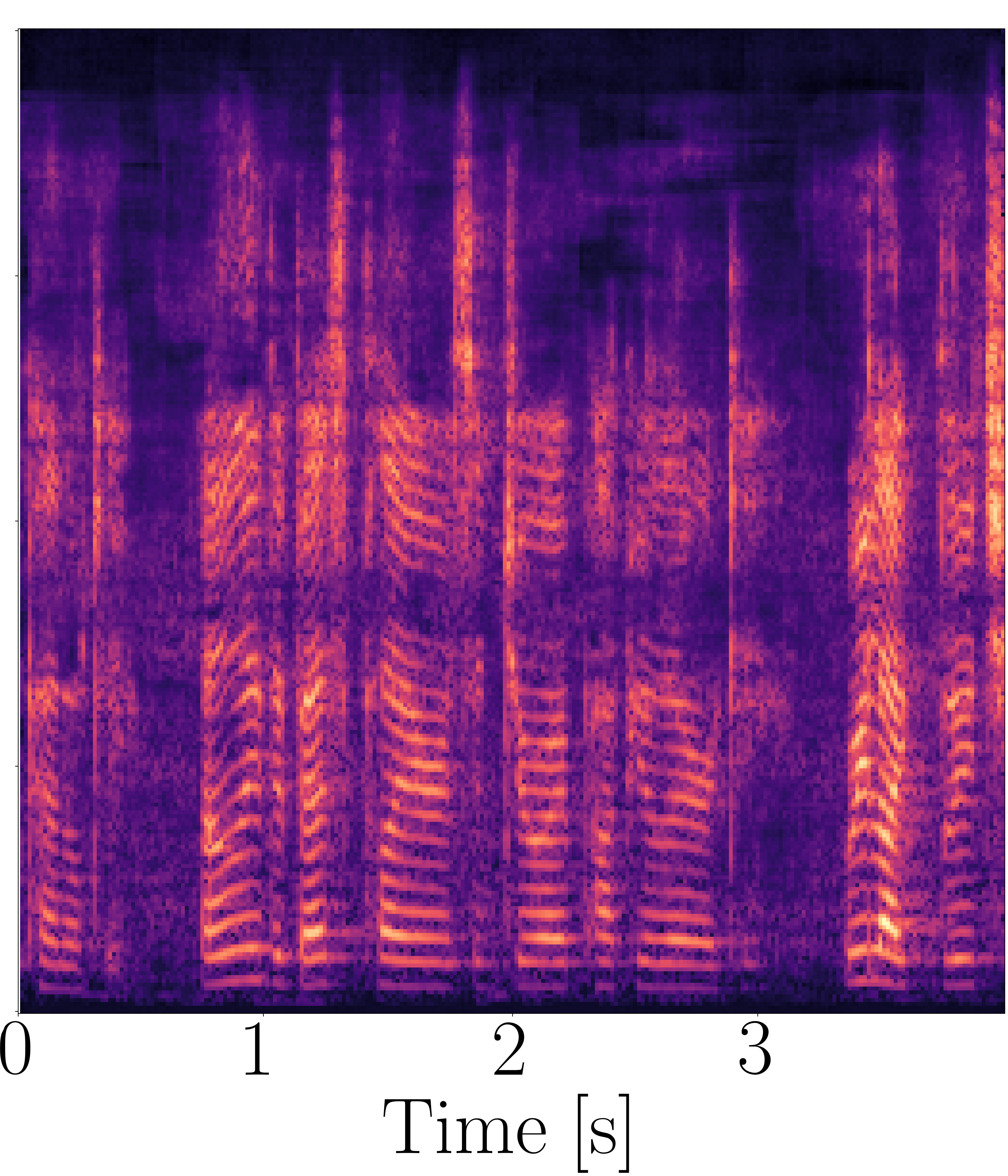}  
          \caption{Denoised.}
          \label{subfig:denoised}
    \end{subfigure}
    \begin{subfigure}{0.31\columnwidth}
          \centering
          \includegraphics[width=1\columnwidth]{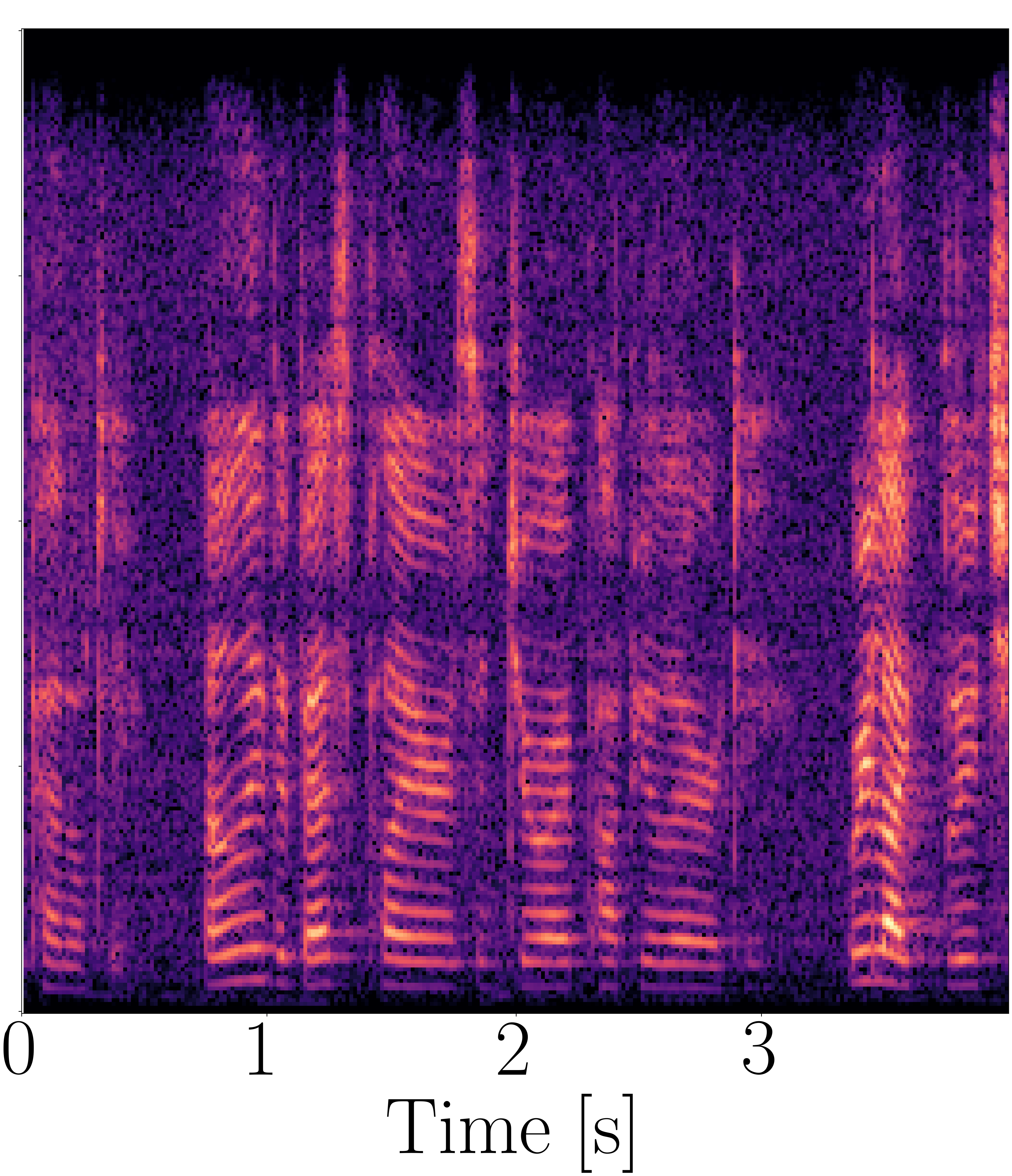}  
          \caption{Original.}
          \label{subfig:original}
    \end{subfigure}
    
    \caption{Example denoised spectrograms obtained by the \gls{dncnn}. The noisy spectrogram in~(\subref{subfig:noisy}) is processed to obtain a denoised version~(\subref{subfig:denoised}). Our goal is to obtain a denoised signal which is as similar as possible to the original one, in~(\subref{subfig:original}).}
    \label{fig:denoising process}
\end{figure}

Fig.~\ref{fig:denoising process} depicts an example of spectrogram denoising for audio with \gls{snr} of $30$dB, in which the network is able to remove the influence of background noise while leaving the intrinsic device traces and the speech component intact.

\subsection{Fingerprint Extraction}
After denoising, we proceed with the extraction of particular fingerprints that aim to model the device characteristics as precisely as possible. If we denote with $\f$ our generic fingerprint, we can write
\begin{equation*}
    \f = \operatorname{FING}(\hat{\PP}),
\end{equation*}
where $\operatorname{FING}(\cdot)$ denotes the specific fingerprint extraction procedure.

In the following, we present and individually explain three alternative procedures commonly used in the literature and devised for this purpose.

\paragraph{$\f_1$ — Channel Response} 
this fingerprint is referable to the multiple tolerances of the hardware components adopted by the different manufacturers in the production process of the device~\cite{hanilci_2012}. Its weakness lies in being sensitive to external signals such as noise and music. Following the procedure in~\cite{gaubitch_2013}, we train a \gls{gmm} architecture on a \gls{rasta} filtered-\gls{mfcc} extracted from a speech-only audio signal. We generate an efficient representation of the speech component with this spectral information, creating a dictionary of spectra associated with language phonemes. Thanks to this estimation, we can generate the speech-only spectrogram through a convex combination of individual spectra of the dictionary. After normalizing this new spectrogram and the denoised one, we perform a spectral subtraction, removing the speech component from the original audio recording and being left with the sole channel influence. 
The resulting residual information ideally represents the data left by the recording device during acquisition. We can estimate the device's channel response by time-averaging the residual spectrogram for each frequency bin.

\paragraph{$\f_2$ — Cuccovillo, 2013} 
this feature, proposed in~\cite{cuccovillo_2013}, involves the union of three different multidimensional components, highlighting different recording's information. The feature extends the channel response described before, embeds information related to the estimated channel, its correlation with the original spectrogram, and information related only to the latter. The authors' goal was to let the feature be independent from the influence of lossy compression and background music.

\paragraph{$\f_3$ — \glsfirst{bed}} 
proposed in~\cite{luo_2018}, this feature follows the assumption that neighboring frequency bands have a high correlation, unlike those distant from each other. This correlation is encoded into a vector based on the values assumed by adjacent frequency bins and corresponds to the time average of the discretized partial frequency derivative of the power spectrum.

\subsection{Classification}
The final block deals with matching the extracted fingerprint and the device model, which we implement through a supervised classifier.
After training the classifier, we estimate the device from a given recording as
\begin{equation*}
    c = \operatorname{CLASS}(\f),
\end{equation*}
with $\operatorname{CLASS}(\cdot)$ indicating the decision rule applied by the classifier and $c\in \mathcal{C}$ the predicted class.
We perform this identification by individually training a classifier for each of the three proposed features, i.e. $\f_1$, $\f_2$, $\f_3$.

\section{Experimental Evaluation}
\label{section:evaluation}
We evaluate the performance of our system by formulating the identification problem in a closed-set scenario, where the possible number of models is limited and known in advance.

\subsection{Datasets}
\label{subsection:dataset}
To carry out the experiments, we train three different architectures, i.e.:
\begin{itemize}
    \item \emph{\glsfirst{svm}} classifier: for the device identification;
    \item \emph{\glsfirst{dncnn}} neural network: for noise removal;
    \item \emph{\glsfirst{gmm}} density estimator: for speech reduction, required by the $\f_1$ and $\f_2$ features.
\end{itemize}

The \gls{svm} classifier is trained with features extracted from the MOBIPHONE dataset~\cite{kotropoulos_2014}. This dataset contains $504$ recordings sampled at $16$ kHz. There are $24$ recordings of $30$ seconds for each of the $21$ mobile phones produced by $7$ different manufacturers. The recordings are sentences spoken by $24$ different speakers, chosen from the TIMIT dataset~\cite{TIMIT_dataset}.
For the \gls{svm} architecture, we provide different training for each feature, and for each one of them, we train three different \glspl{svm}. i.e.:
\begin{itemize}
    \item \emph{Clean}: classical training, with features obtained from noiseless recordings;
    \item \emph{Mixed}: multi-scene data augmentation training, with features obtained from recordings having a variable \gls{snr} between $20$dB and $35$dB and the noiseless one, i.e., the original;
    \item \emph{Denoised}: multi-scene data augmentation training, with features obtained from denoised recordings that had variable \gls{snr} between $20$ and $35$dB and the noiseless one.
\end{itemize}
We train the classifier with a \emph{cross-speaker} strategy to validate the speaker-independent approach further. Specifically, we use $19$ of the $24$ available speakers of the MOBIPHONE dataset to train the classifier and the remaining $5$ to test it. This approach allows us to have a higher degree of confidence in the performance regarding its independence and robustness when testing different speakers from those used in training. The training is performed in a supervised manner, providing the correct feature-device pairs $(\f, c)$ to the classifier. We use a multi-class one-vs-rest \gls{svm} classifier with a \gls{rbf} kernel and $\gamma=1/(L\,\sigma^2_\f)$, where $L$ indicates the adopted feature vector length and $\sigma^2_\f$ is equal to the variance of the normalized feature vectors.

For the \gls{dncnn}, we create a custom version of the MOBIPHONE dataset mentioned above. Our modified dataset involves changing the original time domain to a time-frequency one. We generate Log-Distributed Power Spectrograms with $N=512$ samples window with a $50\%$ overlap, obtaining $M=256$ positive frequency bins, with a sampling rate of $F_\text{s}=16$ kHz. Subsequently, the obtained spectrograms are divided into $M \times M$ non-overlapped patches, resulting in $M$ frequency bins and $M$ time frames. Each spectrogram covers approximately $4$ seconds of the original recording. These settings impose the audio length used for performing the denoising process in our experiments. In addition to this, we generate multiple noisy versions using different \glspl{snr} to simulate a plausible counter-forensics attack scenario with noise injection on the audio recording. We decided to adopt $20$, $25$, $30$, $35$dB \glspl{snr} for a more generalized training.

The neural network is trained with clean/noise pair spectrograms to find a mapping between the two to compute the residual information related to the noisy component, using the Adam optimizer with an initial learning rate of 10e-3. On the aforementioned dataset the training is converging after $64$ epochs. 

The \gls{gmm} density estimator is trained using a one-hour duration speech-only audio track, created by combining recordings from the LibriSpeech corpus~\cite{LibriSpeech_dataset}. This dataset allowed us better accuracy and more generalization compared to the original version in~\cite{gaubitch_2013}.

\subsection{Comparison of Different Features}
This section analyzes the performance of the three different features in noisy conditions and how the denoising process acts to recover the information lost in the view of device identification.

Table~\ref{tab:dncnn effects} shows the different accuracies obtained downstream the identification process with and without the denoiser preprocessing, as the \gls{snr} of the input audio recording varies. A different \gls{svm} classifier was trained for each feature, while the \emph{Clean} (noiseless) training set was kept constant across \glspl{snr}.
Analyzing the results, we can see how the noise removal process improves the performance of all three features.

\begin{table}[hb]
\centering
\caption{Identification Accuracy {[}\%{]} between the three features for different test SNRs, with and without the DnCNN denoiser contribution, using the \emph{Clean} training for the SVM classifier.}
\label{tab:dncnn effects}
\begin{tabular}{@{}ccccccc@{}}
\toprule
\multirow{2}{*}{\vspace{-.8\tabcolsep} Feature} & \multirow{2}{*}{\vspace{-.8\tabcolsep} DnCNN} & \multicolumn{5}{c}{Test Dataset}\\ \cmidrule(l){3-7}
                        & & 20dB & 25dB & 30dB & 35dB & Original \\ \midrule
\multirow{2}{*}{$\f_1$} & \cmark & 57.5 & 69.1 & 83.7 & 84.2 & 95.1 \\ 
                        & \xmark & 36.0 & 41.8 & 50.4 & 60.6 & 99.2 \\ \midrule
\multirow{2}{*}{$\f_2$} & \cmark & 47.2 & 49.3 & 53.9 & 57.5 & 96.5 \\ 
                        & \xmark & 26.7 & 33.3 & 39.7 & 50.8 & 99.9 \\ \midrule
\multirow{2}{*}{$\f_3$} & \cmark & 64.8 & 74.4 & 78.0 & 85.6 & 99.3 \\ 
                        & \xmark & 42.6 & 52.8 & 58.2 & 77.7 & 99.2 \\ \bottomrule
\end{tabular}
\end{table}

The most significant increase is for $\f_1$, since it is based on frequency information and, therefore, very dependent on the action of white Gaussian noise. Thus, the denoising process proves to be essential and suitable for removing most of the signal energy associated with the noise. Regarding $\f_2$, we have a less evident improvement, with nearly constant accuracy for all the test \glspl{snr}. This is justified by the fact that the robustness of this feature is related to the action of compression and environmental noise that have not been examined in this work. $\f_3$, although already quite robust to \gls{awgn} injection, improves a lot its discriminative power, showing an almost stable improvement from $20$ to $30$dB \gls{snr}.

\begin{figure}[ht]
    \captionsetup[subfigure]{aboveskip=0.15em}
    \centering
    \begin{subfigure}{0.5074\columnwidth}
          \centering
          \includegraphics[width=1\columnwidth]{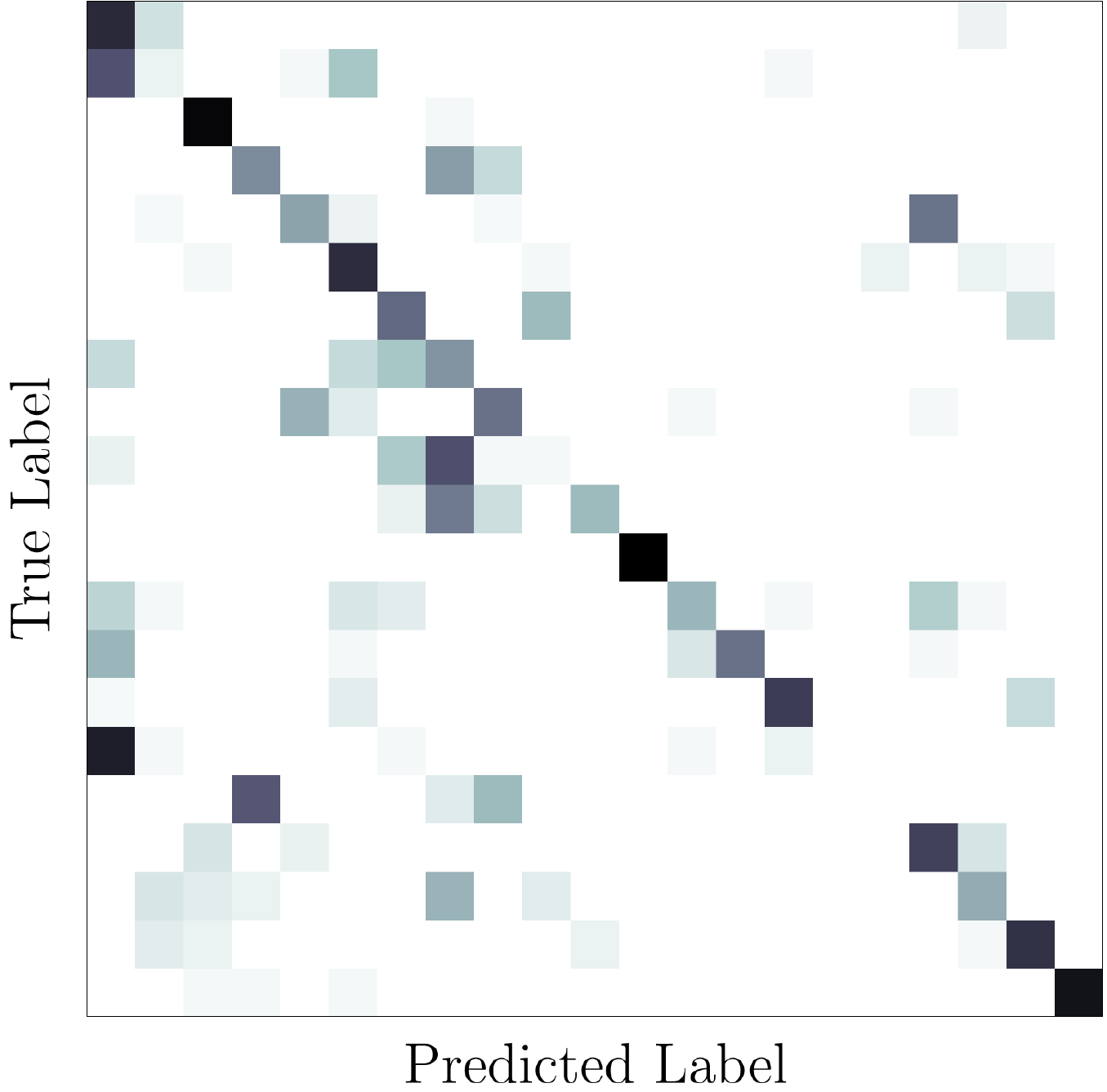}  
          \caption{$\f_1$ without denoising - $50.4$\%}
    \end{subfigure}
    \vspace{1em}
    \begin{subfigure}{0.472\columnwidth}
          \centering
          \includegraphics[width=1\columnwidth]{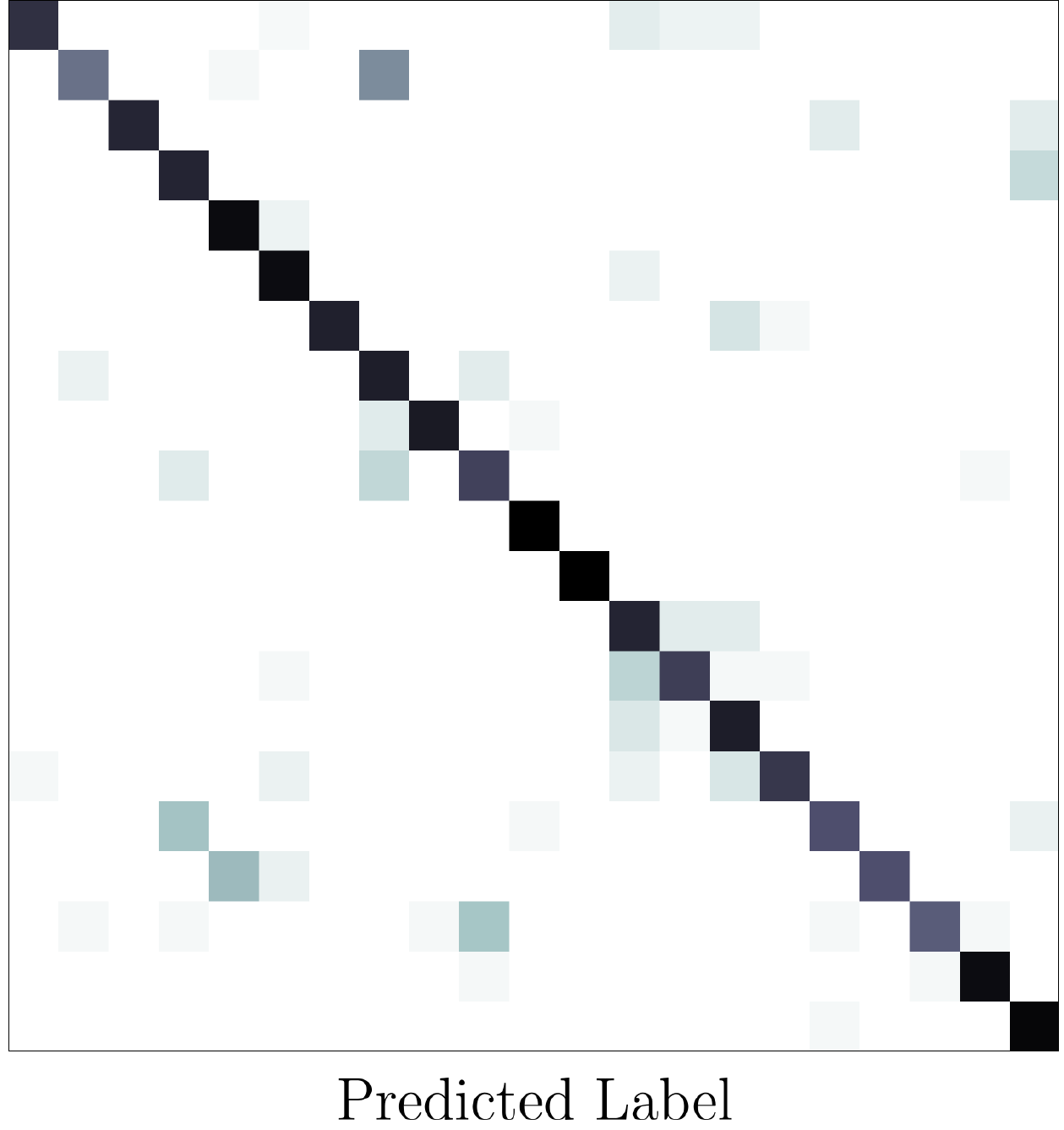}  
          \caption{$\f_1$ with denoising - $83.7$\%}
    \end{subfigure}
    
    \begin{subfigure}{0.5079\columnwidth}
          \centering
          \includegraphics[width=1\columnwidth]{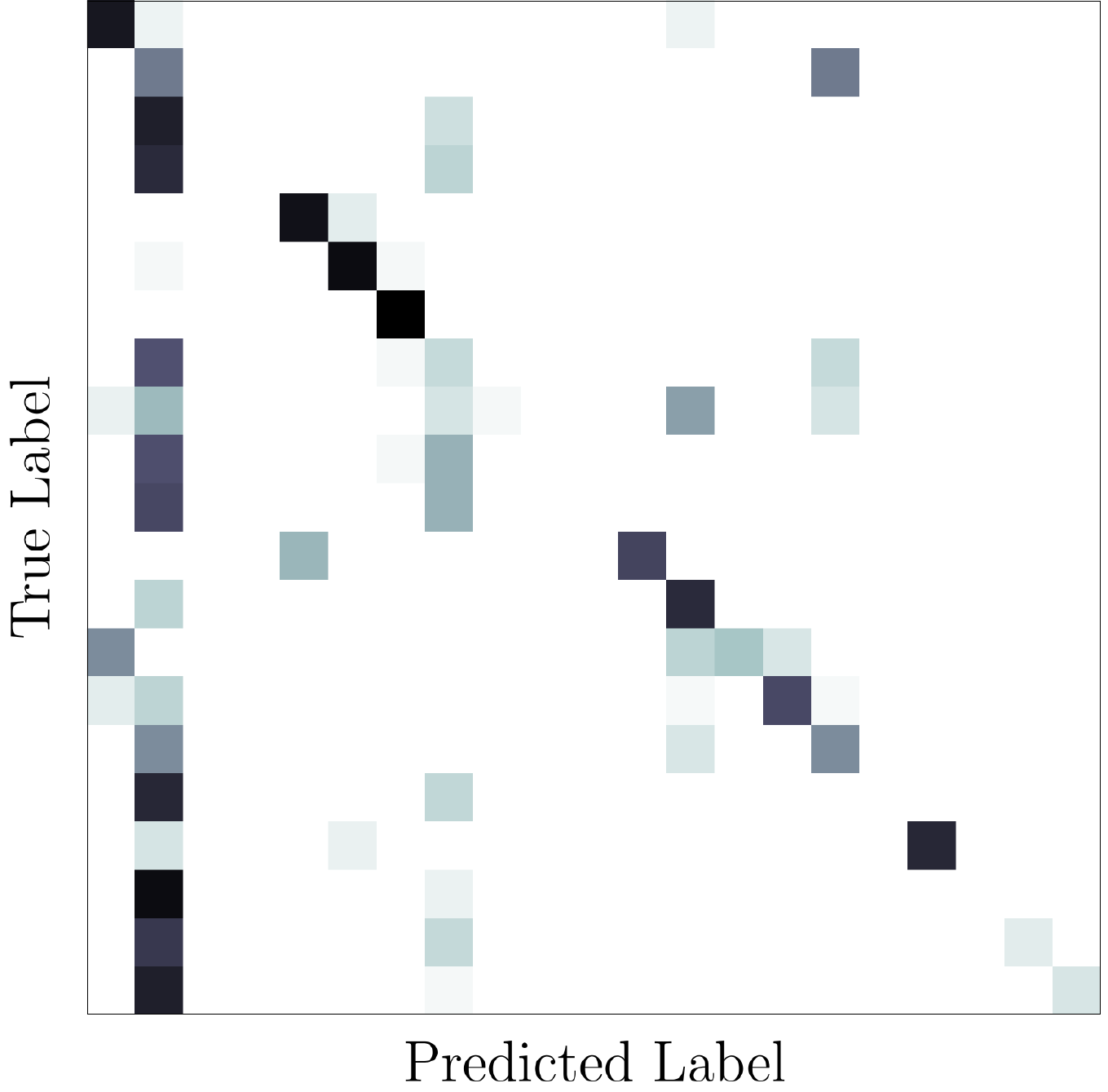}  
          \caption{$\f_2$ without denoising - $39.7$\%}
    \end{subfigure}
    \vspace{1em}
    \begin{subfigure}{0.472\columnwidth}
          \centering
          \includegraphics[width=1\columnwidth]{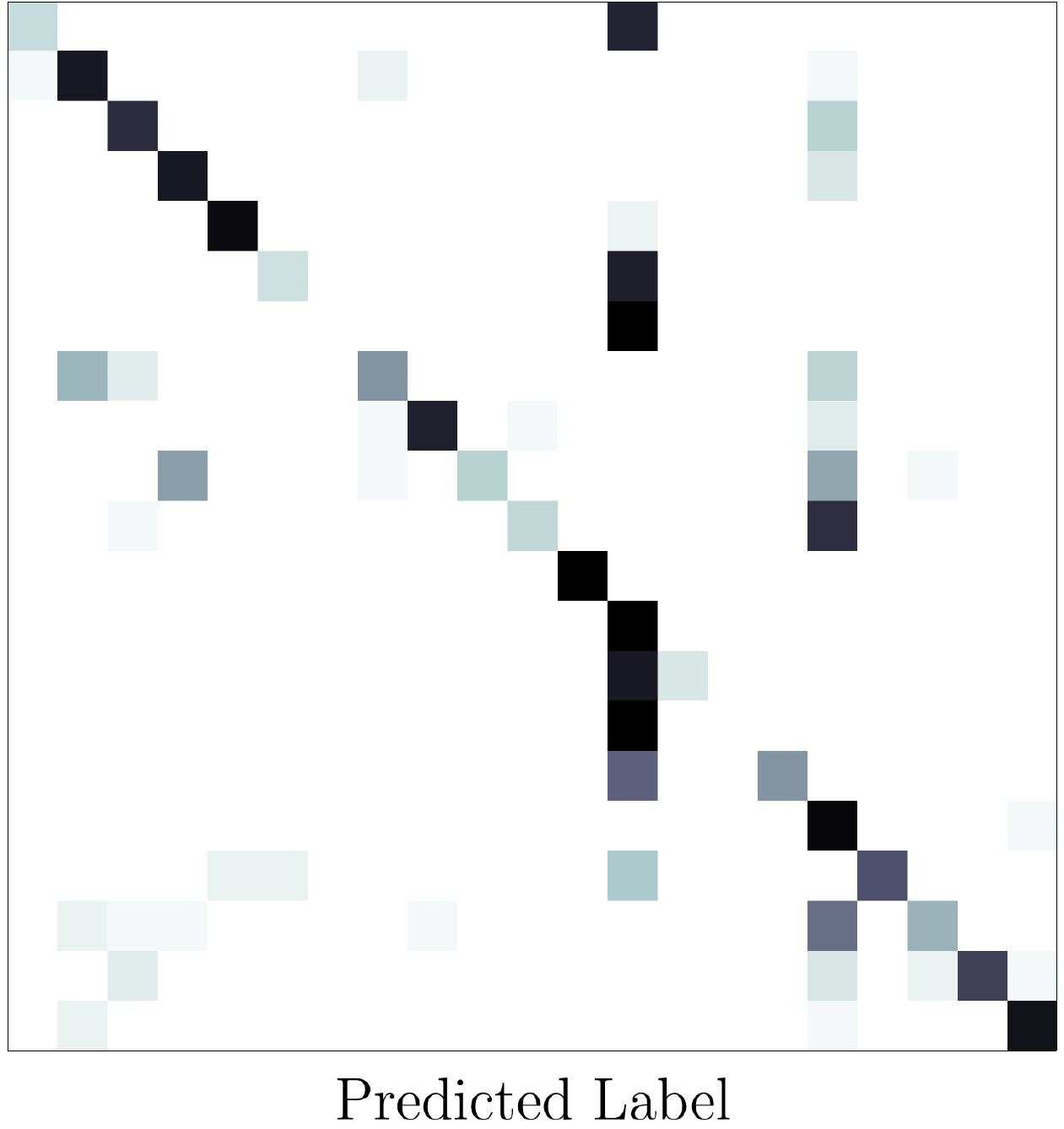}  
          \caption{$\f_2$ with denoising - $53.9$\%}
    \end{subfigure}

    \begin{subfigure}{0.5085\columnwidth}
          \centering
          \includegraphics[width=1\columnwidth]{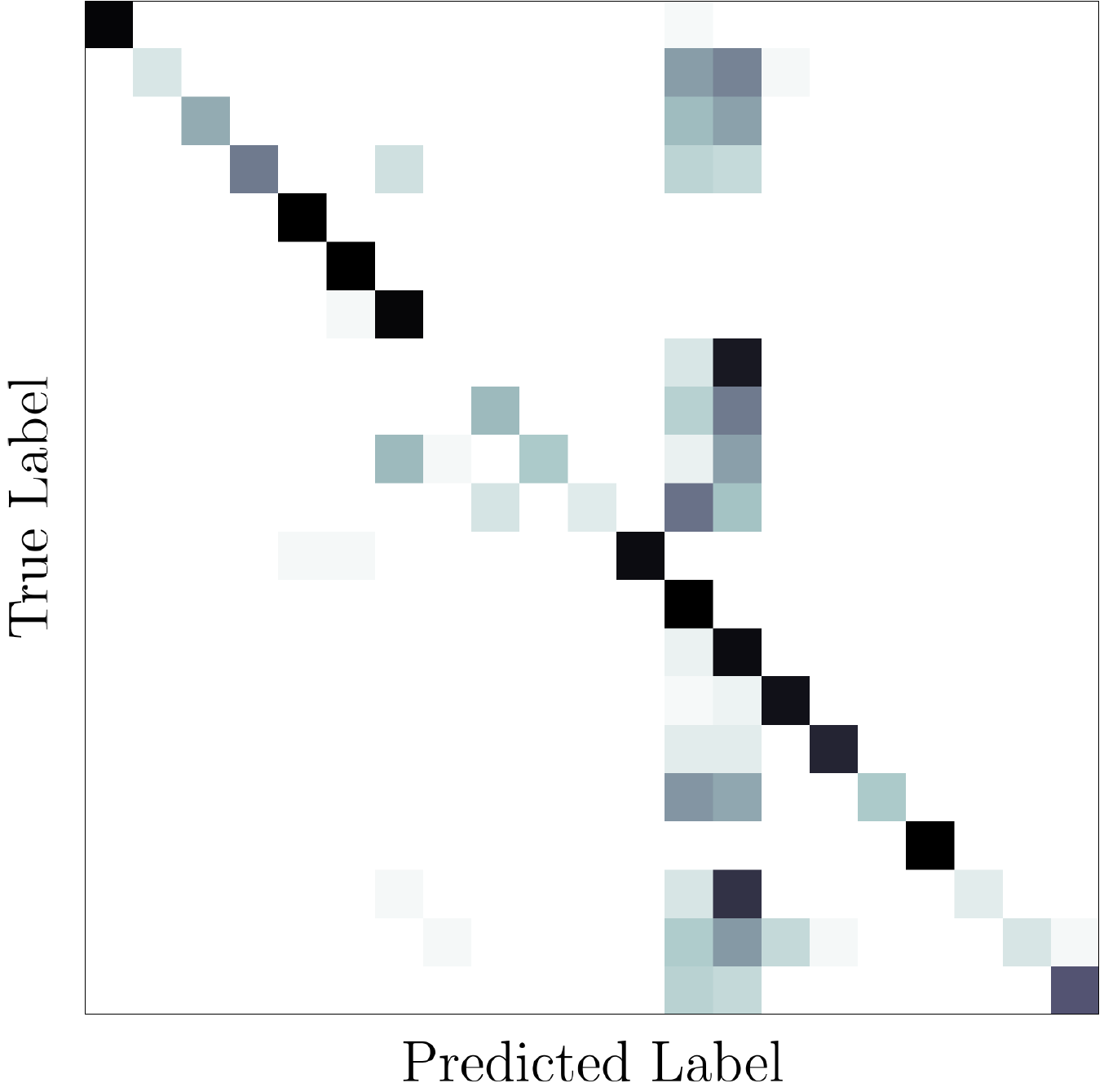}  
          \caption{$\f_3$ without denoising - $58.2$\%}
    \end{subfigure}
    \vspace{1em}
    \begin{subfigure}{0.472\columnwidth}
          \centering
          \includegraphics[width=1\columnwidth]{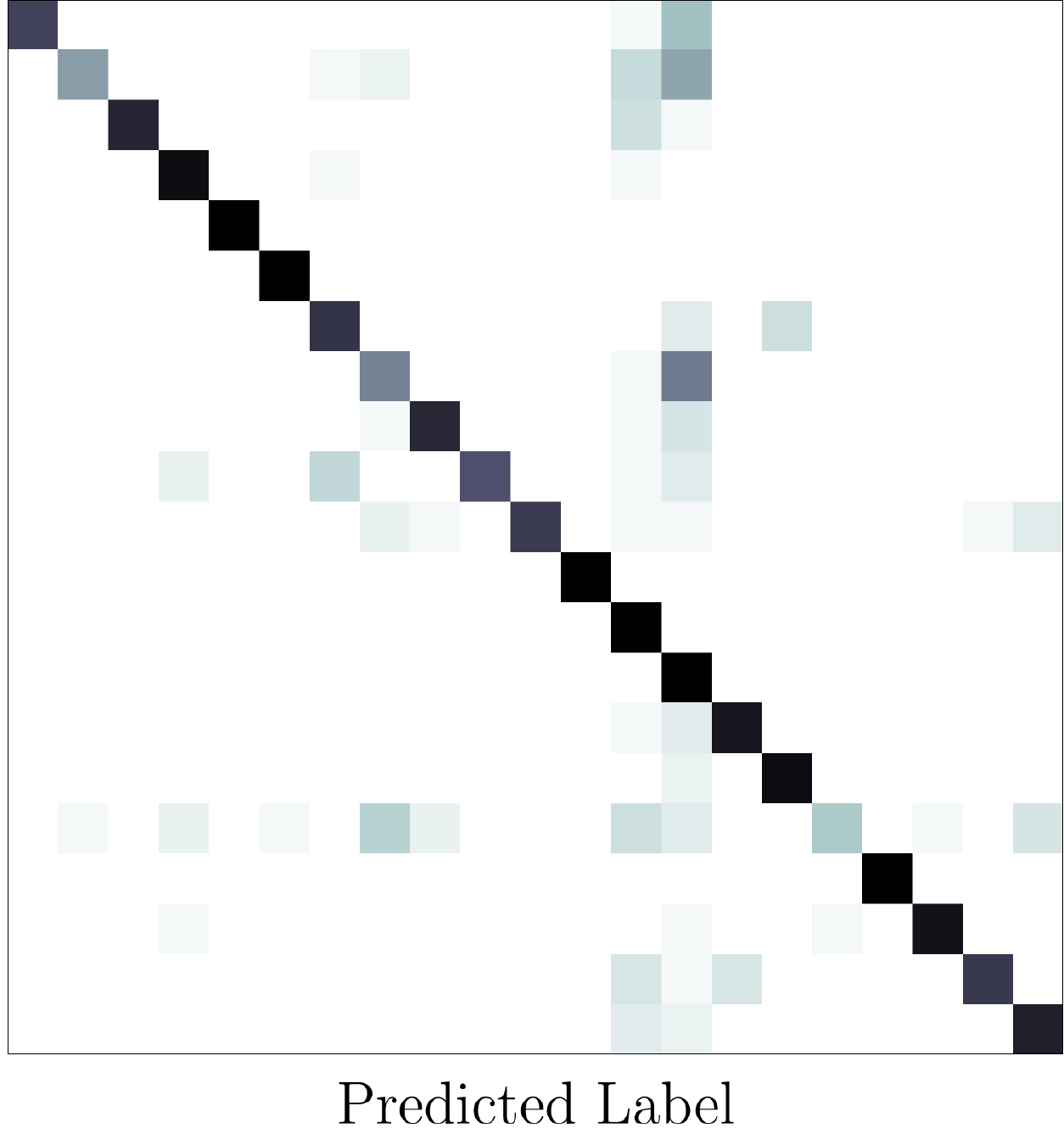}  
          \caption{$\f_3$ with denoising - $78.0$\%}
    \end{subfigure}
    
    \caption{Confusion matrices for all three features, without and with the denoiser preprocessing, for $30$dB SNR recordings and \emph{Clean} SVM training.}
    \label{fig:confusion matrices}
\end{figure}

Fig.~\ref{fig:confusion matrices}, depicts several $J\times J$ confusion matrices for the device identification task on the MOBIPHONE dataset, with the denoiser turned on or off, obtained for $30$dB \gls{snr}. As usual, the matrix element $j_{kl}$ relates to the percentage of elements with predicted class $l$ and having true class $k$, with darker color corresponding to higher accuracy. For the sake of space, we are not reporting the labels of each device in the graph, but they correspond to the MOBIPHONE dataset devices in alphabetical order. By observing the increase in the diagonality of the matrices, we can determine the benefit introduced by the presence of the denoiser for device identification.

\subsection{Training with Data Augmentation}
Up to this point, we tested the system with different \glspl{snr} while using the sole clean files for training. In this section, we want to investigate mitigation strategies from noise corruption instead, testing the system with the three different training of our \gls{svm} classifier, as reported in Section~\ref{subsection:dataset}. The results are shown in Table~\ref{tab:svm different training}: We can observe that M(\emph{ixed}) training turns out to be somehow effective for $\f_2$ and $\f_3$ concerning all five \glspl{snr}, if compared to case C(\emph{lean}). The feature $\f_1$, behaves differently, to the point that from $30$dB \gls{snr} we notice a convenience in training on noiseless recordings. Most importantly, training on the D(\emph{enoised}) set seems to be consistently the best option, leading to a strong boost in the identification performance. Indeed, the results obtained while training on the augmented D(\emph{enoised}) set approach the ones obtained in the noiseless ideal case.

\begin{table}[hb!]
\centering
\caption{Identification Accuracy {[}\%{]} of our system, for all the three features and tested using different SNRs and SVM's training in \ref{subsection:dataset}, i.e., M(ixed), C(lean), D(enoised).}
\label{tab:svm different training}
\resizebox{\columnwidth}{!}{%
\normalsize
\begin{tabular}{@{}cccccccc@{}}
\toprule
\multirow{2}{*}{\vspace{-.8\tabcolsep} Feature} & \multirow{2}{*}{\vspace{-.8\tabcolsep} DnCNN} & \multirow{2}{*}{\vspace{-.8\tabcolsep} \splitcell[c]{Training\\ Dataset}} &\multicolumn{5}{c}{Test Dataset}\\ \cmidrule(l){4-8}
                        &        & & 20dB & 25dB & 30dB & 35dB & Original \\ \midrule
\multirow{3}{*}{$\f_1$} & \xmark & M & 67.3 & 68.9 & 67.3 & 68.0 & 83.4 \\ 
                        & \cmark & C & 57.5 & 69.1 & 83.7 & 84.2 & 95.1 \\
                        & \cmark & D & 83.6 & 92.6 & 96.7 & 95.6 & 95.3 \\ \midrule 
\multirow{3}{*}{$\f_2$} & \xmark & M & 86.0 & 94.2 & 96.3 & 96.2 & 94.3 \\ 
                        & \cmark & C & 47.2 & 49.3 & 53.9 & 57.5 & 96.5 \\
                        & \cmark & D & 92.7 & 96.2 & 98.1 & 98.5 & 93.5 \\ \midrule 
\multirow{3}{*}{$\f_3$} & \xmark & M & 75.3 & 85.6 & 91.9 & 96.5 & 98.5 \\ 
                        & \cmark & C & 64.8 & 74.4 & 78.0 & 85.6 & 99.3 \\
                        & \cmark & D & 93.6 & 97.8 & 98.4 & 98.0 & 96.2 \\ \bottomrule
\end{tabular}
}
\end{table}

\section{Conclusions}
In this paper, we propose a robust technique for speaker-independent microphone identification under noisy conditions. In particular, our study investigated a new methodology based on a residual denoiser acting on the time-frequency domain of the audio speech recording. Its goal is to remove unnecessary information related to the unwanted noise, leaving only the signal parts that characterize the device allowing it to be unequivocally distinguished.

The proposed method outperforms the state-of-the-art on the \gls{awgn}-corrupted MOBIPHONE dataset and was also tested with different speakers used in the training phase, thus solving the problem in a speaker-independent manner. In addition to our previous work in \cite{cuccovillo_2022_arXiv}, we also determined that the method is applicable to three different features for microphone classification, and we observed that data-driven denoising at both training and testing phase is required for a significant increase in performance.

Future developments will involve developing new techniques for speech reduction addressing sex, age, and language generalization. Two further aspects that need further analysis are robustness against audio compression and the effect of different noise types. Furthermore, there is a need for larger datasets including a broad range of devices and recording conditions. Currently, such datasets are not available in the research community. Lastly, it could be possible to replace the feature extraction or the classification stage with an ad-hoc neural network.

\section*{Acknowledgment}
This paper was supported by the EU H2020 AI4Media research project (grant no. 951911).

\bibliographystyle{IEEEtran}
\bibliography{ref}
\vspace{12pt}
\end{document}